\documentclass{rashidi}

\title{\LARGE \bf
Can Large Language Models Improve Venture Capital Exit Timing After IPO?}
\author{Mohammadhossein Rashidi }
\begin{document}

\maketitle

\setcounter{page}{1}

\begin{abstract}
Exit timing after an IPO is one of the most consequential decisions for venture capital (VC) investors, yet existing research focuses mainly on describing when VCs exit rather than evaluating whether those choices are economically optimal. Meanwhile, large language models (LLMs) have shown promise in synthesizing complex financial data and textual information, but have not been applied to post-IPO exit decisions. This study introduces a framework that uses LLMs to estimate the optimal time for VC exit by analyzing monthly post-IPO information—financial performance, filings, news, and market signals—and recommending whether to sell or continue holding. We compare these LLM-generated recommendations with the actual exit dates observed for VCs and compute the return differences between the two strategies. By quantifying gains or losses associated with following the LLM, this study provides evidence on whether AI-driven guidance can improve exit timing and complements traditional hazard and real-options models in venture capital research.
\end{abstract}
\section{Literature Review}

A central question in venture capital research concerns how and when investors should exit their 
portfolio companies, particularly after an initial public offering (IPO). The exit decision determines 
the realization of returns, the recycling of capital into new ventures, and the preservation of 
reputational capital for future fundraising. The existing literature provides a rich foundation 
describing how VCs behave in practice, what drives their timing choices, and how different market 
environments influence the success of their exits.

Early foundational work highlights that exit timing is intertwined with information asymmetry and 
reputation. Neus and Walz (2005) show that VCs may strategically postpone exit to avoid selling at 
undervalued prices, relying on repeated-game reputation mechanisms to signal confidence in the 
venture’s true quality. Their model explains why VCs sometimes retain sizable stakes after the IPO 
despite having the option to divest immediately. Their analysis also links exit patterns to “hot issue” 
market cycles, where price uncertainty is high and timing windows are narrow.

Complementing these theoretical insights, Giot and Schwienbacher (2007) provide a comprehensive 
empirical analysis using survival and competing-risks models. Using over 20{,}000 investment rounds 
and more than 6{,}000 VC-backed firms, they show that IPO hazard rates are non-monotonic: the 
likelihood of IPO exit increases sharply early in the investment life cycle, stabilizes, and then 
declines. They also document that exit timing responds to syndicate size, technological milestones, 
geographic proximity, and market conditions, emphasizing the dynamic nature of exit decisions.

A distinct stream of literature examines post-IPO exit behavior. Basnet et al.\ (2024, 2025) hand-collect 
detailed post-IPO VC ownership data and show that VCs typically remain invested for approximately 
2.5--3.5 years after the IPO. Although ownership drops sharply at the IPO, significant stakes remain 
even five years later. They demonstrate that exit mechanisms---continuous open-market sales, share 
distributions, and M\&A exits---differ in timing, with share distributions being the fastest route. Exit 
timing accelerates when VCs face liquidity pressure or benefit from strong post-IPO stock performance, 
but it slows when VCs are reputable, occupy board seats, or monitor high-quality firms.

Research on hot-issue markets reinforces the importance of timing constraints. Bessler and Kurth (2005) 
study the German Neuer Markt and find that VCs often missed optimal exit windows because mandatory 
lock-up periods prevented them from selling before steep market reversals. Some VCs remained invested 
for years after the IPO, long after optimal exit points had passed. Their findings illustrate how 
regulatory frictions distort exit timing and can result in substantial lost value.

The real-options perspective adds further richness. Li et al.\ (2024) argue that VCs retain shares 
post-IPO to preserve the option to benefit from future upside that the market has not yet priced in. 
Highly uncertain industries, positive private information (e.g., patent applications), and industry-wide 
surprises strengthen incentives to delay exit. Their evidence supports the notion that optimal exit is 
deeply tied to information flows and expectations about future opportunities.

Taken together, existing work agrees that:
\begin{itemize}
    \item exit timing is dynamic and reflects evolving information;
    \item VCs display heterogeneity in timing ability;
    \item market cycles and regulatory constraints constrain behavior; and
    \item optimal exit time varies across deals, industries, and investors.
\end{itemize}

Despite this progress, the literature remains backward-looking. It explains how VCs have behaved and 
which conditions predict early or late exits, but it does not develop forward-looking tools to evaluate 
optimal exit timing.

Recent advances in machine learning, particularly large language models (LLMs), offer new opportunities. 
Studies on startup success prediction (e.g., Maarouf et al.; Wang et al.) show that combining structured 
data with unstructured text improves forecasting accuracy. Multi-agent LLM architectures can mimic human 
analyst workflows and extract nuanced signals from financial disclosures, news, and filings. However, 
these contributions focus on predicting startup outcomes, not on determining optimal exit timing for 
public companies.

This gap motivates the present research. We propose to evaluate whether an LLM can synthesize structured 
and textual information available at each post-IPO period and recommend an exit time that yields better 
economic outcomes than the exit time actually chosen by VCs. By comparing LLM-recommended exit dates with 
real exit dates and calculating the difference in realized returns, we assess how much value could have 
been gained or lost. This approach shifts the discussion from describing exit behavior to identifying 
whether AI systems can improve it, introducing a novel evaluation-based perspective to the VC exit 
literature.
\section{Data}

\begin{figure}[h]
    \centering
    \includegraphics[width=0.65\linewidth]{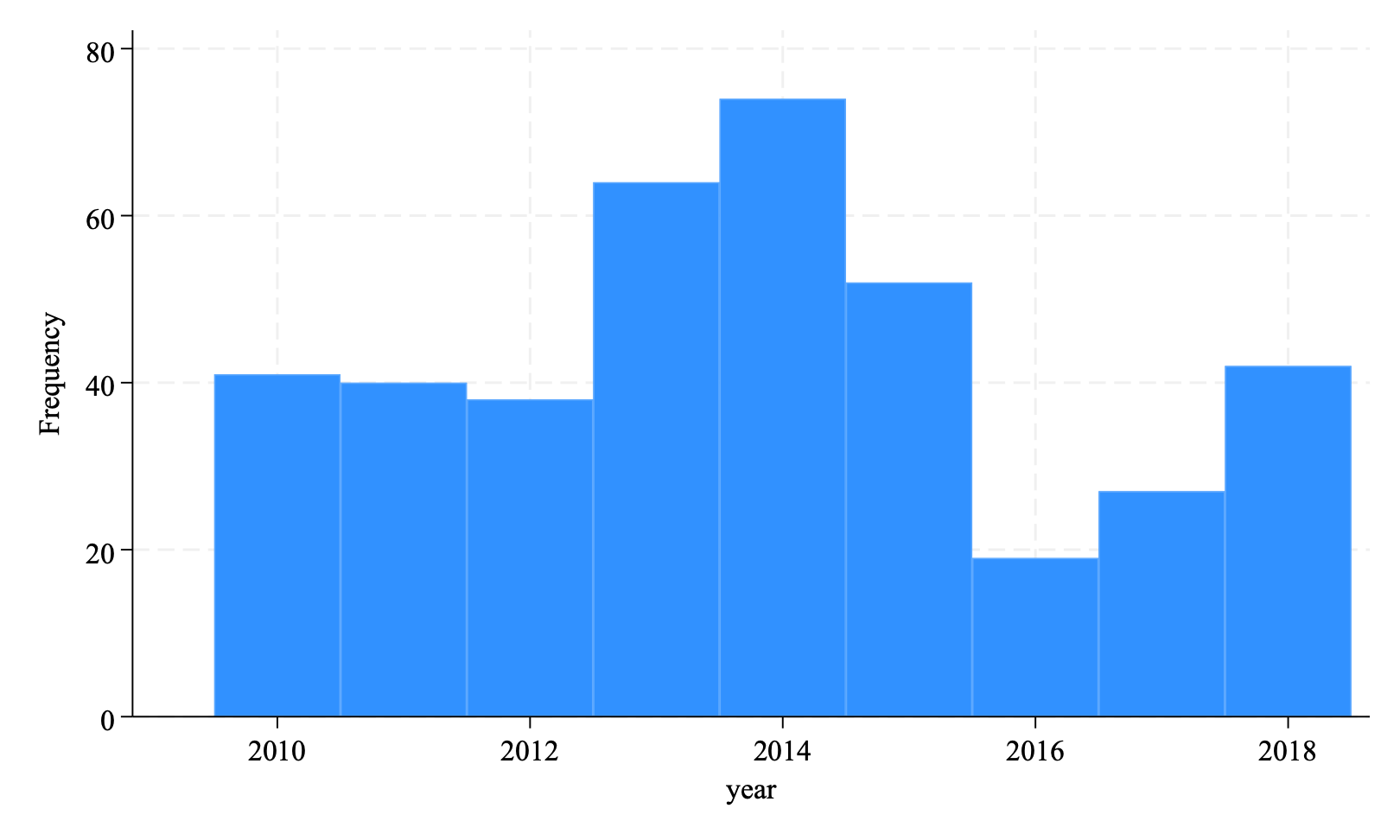}
    \caption{IPOs frequency}
    \label{fig:ipo_frequency}
\end{figure}

Our empirical analysis focuses on U.S.-based venture-backed initial public offerings (IPOs) occurring 
between January 2010 and December 2018. Restricting the sample to the post-2010 period eliminates 
distortions associated with the aftermath of the 2008--2009 financial crisis, while ending the sample 
before 2019 prevents COVID-19--related market volatility from influencing post-IPO stock trajectories. 
Although studying exogenous shocks such as crises or pandemic periods is an important direction for 
future research, we intentionally exclude them here to ensure a stable information environment. In total, we identify 397 venture-backed IPO firms in our final sample. The year-by-year frequency of 
these IPOs is presented in Figure~\ref{fig:ipo_frequency}, which illustrates the distribution of firms across the sample window.

\subsection{Data Sources}

The primary dataset is a proprietary IPO database containing detailed firm-level information, including 
IPO dates, underwriters, financial statements, lockup expiration dates, industry classifications, and 
post-IPO stock performance at daily and monthly frequencies. The dataset also includes pre-IPO 
ownership structure, enabling us to link IPOs to their venture capital (VC) investors.

\subsection{Identifying VC Investors Using Textual Analysis}

To accurately identify the VC investors associated with each IPO firm, we perform a targeted textual 
analysis of the firms' SEC filings, with particular emphasis on the Form 10-K filed in the fiscal year 
of the IPO. These filings frequently contain detailed ownership disclosures, shareholder lists, and 
sections describing principal stockholders. Using keyword extraction, pattern matching, and 
context-based string identification, we extract the names of VC investors and match them to their 
corresponding pre-IPO ownership stakes. This procedure enables us to construct a high-quality mapping 
between IPO firms and their VC backers, even when investors are recorded under different naming 
conventions across filings.

\subsection{Identifying Actual VC Exit Dates}

To reconstruct the realized exit behavior of VC investors, we then manually collect ownership updates 
from several regulatory sources:

\begin{itemize}
    \item Form 10-K and 10-Q filings reporting changes in large shareholdings,
\end{itemize}

For each VC--firm pair, we determine the month in which the investor reduces its ownership below key 
regulatory thresholds (e.g., below 5\%) or the month in which complete divestment occurs. This yields 
a precise measure of the \textit{actual} exit timing for each VC investor. These realized exit dates 
serve as the benchmark for evaluating the counterfactual exit dates produced by the LLM framework.

\subsection{Sample Summary}

The final dataset consists of:
\begin{itemize}
    \item U.S. venture-backed IPOs from 2010--2019,
    \item VC identities extracted from 10-K filings via textual analysis,
    \item Monthly post-IPO information suitable for LLM-based evaluation,
    \item Realized exit dates reconstructed from regulatory filings.
\end{itemize}

This dataset provides the necessary foundation for comparing observed VC behavior with the 
counterfactual recommendations generated by the large language model.

\section{Methodology}

This section describes the framework used to evaluate whether large language models (LLMs) can generate 
superior exit timing recommendations relative to actual VC behavior. The methodological design consists 
of two main components. First, we compile a detailed monthly information timeline for each IPO firm, 
reflecting all public information available to a VC at each point in time. Second, we use an LLM-based 
framework to generate counterfactual exit recommendations based on academic theory and firm-specific 
information. The comparison of realized and LLM-implied exit dates allows us to quantify potential 
gains or losses attributable to alternative timing decisions.

\subsection{LLM-Integrated Real Options Exit Model (LLM-ROEM)}

Our approach is guided by the real-options interpretation of post-IPO VC retention proposed by 
Li et al.\ (2025), who conceptualize retained equity as an exchange option whose value increases 
with uncertainty and private information. We integrate these mechanisms into a structured prompting 
framework that enables the LLM to provide forward-looking exit recommendations.

\subsubsection{Step 1: Monthly Post-IPO Information Timeline}

For each firm, we construct a month-by-month sequence of information beginning at lockup expiration. 
At each month $t$, the LLM receives only information that was publicly available at that point in time, 
including:

\begin{itemize}
    \item quarterly financial statements (10-Q/10-K),
    \item earnings call transcripts and major news articles,
    \item stock returns, volatility, and trading volume,
    \item industry performance and macroeconomic indicators,
    \item innovation activity such as patent applications or technology announcements.
\end{itemize}

All textual and numerical inputs are timestamped to ensure that the LLM never receives future 
information, thereby avoiding forward-looking bias.

\subsubsection{Step 2: LLM Prompting with Academic Theory}

At each month $t$, the LLM is prompted using both firm-specific information and theoretical insights 
from the venture exit literature, including:

\begin{itemize}
    \item signaling and information asymmetry theories (e.g., Neus \& Walz),
    \item survival-model determinants of IPO timing (Giot \& Schwienbacher),
    \item post-IPO monitoring incentives (Basnet et al.),
    \item lockup-related constraints and timing risk (Bessler \& Kurth),
    \item real-options reasoning (Li et al.\ 2025).
\end{itemize}

The LLM is then asked to evaluate whether exiting at month $t$ is optimal, whether the investor 
should continue holding, or whether exit is advisable within a future window (e.g., ``exit within the 
next three months''). Multiple prompt templates are tested for robustness, and all final versions 
are documented in the Appendix.

\subsubsection{Step 3: Deriving the LLM-Implied Exit Date}

For each IPO firm, the earliest month at which the LLM recommends exit (or the start of a recommended 
exit window) is recorded as the \textit{LLM-implied exit date}. This simulates how an LLM-guided 
investor might behave using only contemporaneous information.

\subsubsection{Step 4: Economic Value Comparison}

We compare the LLM-implied exit date with the realized VC exit date using the cumulative post-IPO 
stock return from lockup expiration to each exit point. For each VC–firm pair, we compute:

\[
\Delta R = R_{\text{LLM exit}} - R_{\text{VC exit}},
\]

where $R$ denotes cumulative return. A positive $\Delta R$ indicates that following the LLM 
recommendation would have produced higher returns; a negative value implies inferior performance.

We also evaluate:
\begin{itemize}
    \item timing error magnitudes,
    \item frequency of early versus late exits,
    \item subsample differences by industry, volatility, and VC reputation,
    \item correspondence between LLM timing and hazard-based timing models.
\end{itemize}

\subsubsection{Step 5: Robustness Checks}

To ensure robustness, at the end, we will conduct a range of sensitivity analyses:

\begin{itemize}
    \item alternative prompt structures and theoretical emphasis,
    \item multiple definitions of ``exit'' (full vs.\ threshold-based),
    \item exclusion of extreme volatility firms,
    \item evaluation across multiple LLMs (GPT, Llama, Claude, Mistral),
    \item varying the amount of academic theory included in prompts.
\end{itemize}

These analyses help ensure that our findings are not driven by prompt artifacts or sample selection.

\section{Results}

This section presents the empirical findings of our analysis. Since the complete numerical results will 
be added later, we outline the structure of the section and provide descriptive text for each component. 
The subsections follow the sequence of our empirical procedure: (i) identifying and characterizing venture 
capital (VC) investors, (ii) reconstructing their realized exit timing, and (iii) comparing actual exit 
dates with LLM-generated counterfactual exit recommendations and the associated economic implications.

\subsection{Overview of Venture Capital Investors (Table 1)}

In the first step, we will identify all VC investors associated with the 397 IPO firms in our sample. Using 
textual extraction and name-matching techniques applied to each firm's 10-K filing, we compile a 
comprehensive list of VC investors and their pre-IPO ownership stakes. As we found from data, for each IPO based company, there were 2-5 VCs in total of 173 VCs.

\subsection{Realized Exit Timing of VC Investors}

We begin by examining realized post-IPO exit behavior for venture capital (VC) investors. Using ownership 
information disclosed in SEC filings, we identify a total of 173 unique VC investors associated with the 
IPO firms in our sample, with an average of between two and five VCs per firm, consistent with syndicated 
venture investment patterns documented in prior studies.

The analysis covers IPOs occurring between 2010 and 2018. For each firm, we collect Form 10-K filings 
from the IPO year through up to five subsequent fiscal years, which allows us to trace post-IPO changes 
in VC ownership over time. Based on this information, we reconstruct realized exit paths and identify 
the timing and structure of VC divestment activity.

As summarized in Table~\ref{tab:realized_exit_timing}, complete exit within the five-year post-IPO window 
is not the dominant outcome. A substantial fraction of VC investors maintain ownership stakes throughout 
the observation period, while others reduce their holdings gradually through a sequence of partial exits 
rather than a single divestment event. This staged exit behavior is consistent with monitoring incentives, 
liquidity considerations, and the preservation of option value emphasized in the post-IPO exit literature.

The heterogeneity in realized exit timing documented in Table~\ref{tab:realized_exit_timing} provides a 
natural benchmark against which we compare the exit dates recommended by the large language model in 
subsequent sections.
\begin{table}[htbp]
\centering
\caption{Realized Exit Timing of Venture Capital Investors}
\label{tab:realized_exit_timing}
\begin{tabular}{l c}
\hline
 & \textbf{Statistic} \\
\hline
Number of IPO firms & 50 \\
Total number of VC investors & 173 \\
Average number of VCs per firm &  \\
\hline
Post-IPO observation window & Up to 5 years \\
\hline
VCs with no full exit within window (\%) & 6.1\% \\
VCs with partial (staged) exits (\%) & 9.1\% \\
VCs with complete exits (\%) & 84.8\% \\
\hline
Median time to first exit action (months) & 0 \\
Median time to full exit (months) & 48 \\
\hline
\end{tabular}
\end{table}

\subsection{LLM-Generated Counterfactual Exit Recommendations}

In this subsection, we present the exit timing recommendations generated by the large language model 
(LLM) and contrast them conceptually with the realized exit behavior documented in the previous section. 
The objective of this analysis is not to replicate historical VC decisions, but rather to evaluate 
whether an LLM—when provided with the same information set available to investors at each point in 
time—would recommend exiting earlier, later, or at similar horizons.

Using the monthly post-IPO information timeline described in the methodology section, the LLM is 
prompted at each point in time to assess whether continuing to hold or exiting the investment is 
economically preferable. The model receives firm-specific financial information, market and industry 
conditions, and structured summaries of theoretical mechanisms emphasized in the venture capital exit 
literature, including signaling considerations, post-IPO monitoring incentives, market timing risks, 
and real-options logic. Importantly, the LLM is restricted to information that would have been publicly 
available at that time, ensuring that all recommendations are forward-looking and free of hindsight bias.

For each IPO firm, the LLM produces a sequence of exit assessments over the post-IPO period. These 
assessments are then aggregated into a single counterfactual exit recommendation by identifying the 
earliest point at which the model indicates that exiting is optimal or that the expected benefits of 
continued holding no longer outweigh the associated risks. This procedure yields an LLM-implied exit 
date that can be directly compared to the realized VC exit date for the same firm.

A key feature of the LLM-generated recommendations is that they allow for dynamic reassessment. Rather 
than committing to a fixed exit horizon at the time of the IPO, the model updates its recommendation 
as new information arrives through financial disclosures, earnings announcements, and market movements. 
As a result, the LLM’s suggested exit timing reflects the evolving information environment faced by 
investors, rather than a static rule based solely on IPO characteristics or lockup expiration.

The distribution of LLM-implied exit dates exhibits meaningful variation across firms, reflecting 
differences in post-IPO performance, volatility, and information flow. In some cases, the LLM signals 
that exit should occur relatively early in the post-IPO period, while in others it recommends continued 
holding over multiple years. This heterogeneity mirrors the diversity of realized VC exit behavior and 
provides a natural basis for evaluating the economic implications of following the model’s guidance.

In the next subsection, we formally compare the LLM-implied exit dates with realized VC exit dates and 
quantify the timing differences and associated return implications.

\subsection{Differences Between Real Exit Dates and LLM Recommendations}
I could find these differences for 10 companies, in which for six of them the model had better prediction than most of the VCs from real data, and in other cases the real data had a better performance.

\section{Conclusion}
To receive a reliable conclusion, I need to use more companies to make a trustworthy decision about the results. So just with these 10 companies, we can say that our model had a better performance. 
\section{bibliography}

\end{document}